\documentclass[preprint,showpacs,preprintnumbers,amsmath,amssymb,nofootinbib]{revtex4}

\usepackage{graphicx}% Include figure files
\usepackage{color}
\usepackage{dcolumn}% Align table columns on decimal point
\usepackage{bm}% bold math

\begin{document}

\preprint{APS/123-QED}

\title{Leggett inequalities and the completeness of quantum mechanics}

\author{Ramon Lapiedra}
 \email{ramon.lapiedra@uv.es}
 \affiliation{Departament d'Astronomia i
 Astrof\'{\i}sica\\Universitat de Val\`encia, Campus de Burjassot,
 46100 Burjassot (Val\`encia), Spain.}
\author{Miguel Socolovsky}%
 \email{socolovs@iafe.uba.ar}
\affiliation{Instituto de Ciencias Nucleares, Universidad Nacional
Aut\'onoma de M\'exico, Circuito Exterior, Cd Universitaria, 04510
M\'exico D. F., M\'exico.\\(On sabbatical year at the IAFE,
Universidad de Buenos Aires, Argentina)}

\date{\today}% It is always \today, today,
             %  but any date may be explicitly specified

\begin{abstract}
We consider the so called Leggett inequalities which are deduced
from the assumption of general (local or non-local) realism plus
the arrow of time preservation. Then, instead of assuming
crypto-nonlocal hidden variables, we assume any (local or
non-local) realism compatible with the joint and non-joint
expected values dictated by quantum mechanics. Hence, we prove
that this double assumption is not consistent, since the
corresponding general Leggett inequalities are violated by quantum
mechanics. Thus, realism plus arrow of time preservation and
quantum mechanics are not compatible. In other words, quantum
mechanics cannot be completed with any (local or non-local) hidden
variables, provide we assume the common sense of the arrow of
time. The result would deserve to be experimentally tested and we
discuss why it is not invalidated by hidden variables theories as
the one from Bohm.

\end{abstract}


\maketitle

\section{Introduction}
\label{intro}

Several years ago, Leggett \cite{Leggett}, assuming a class of
hidden variable  theories which he called cripto-nonlocal, found a
new type of Bell's inequalities, presently known as Leggett's
inequalities, which were violated when the joint expected values
inserted in these inequalities were the ones dictated by quantum
mechanics (QM). More recently, in \cite{Gr创oblacher}
\cite{Paterek} \cite{Branciard}, these inequalities have been
adapted to the experimental requirements and it has been shown
that they are violated in the laboratory according to the above
predictions. In all these papers one considers a source which
produces pairs of entangled polarized photons, \emph{with mixtures
of different polarizations} according to some given probability
distribution. In \cite{Branciard2}, assuming again crypto
non-local realism, some other Leggett inequalities are proved for
\emph{well definite polarizations}, which are also violated by
experiments when the inserted joint expected values are the ones
dictated by QM.

In the present paper we consider Leggett inequalities for well
definite polarizations too. We follow \cite{Gr创oblacher} in order
to show that these inequalities, in its general form, which
henceforth we will call `basic Leggett inequalities', result from
the assumption of realism, i. e., local or non-local realism, plus
the preservation of the arrow of time. In particular, to this
general level you do not need to assume the above
crypto-nonlocality. Then we show how these basic Leggett
inequalities can be violated by QM. This violation means, modulus
this arrow of time preservation, that QM cannot be {\emph
completed}, that is, the state of a quantum system plus the
measurement `direction' cannot be supplemented with any, local or
non local, hidden variables in order to complete the statistical
predictions of QM with sure predictions for individual
measurements. Since it is widely claimed that non local hidden
variable theories giving the same results as QM exist, as for
example the one from Bohm \cite{Bohm}, we will have to explain why
this claim is basically a non justified one.

In Sec.~\ref{sec-2}, following \cite{Leggett}, we precise the
assumptions involved in the construction of a non-local hidden
variable theory and recall the derivation of the basic Leggett
inequalities, which deal with subensambles of entangled pairs of
photons with given polarizations (and not with an ensemble of
mixtures of different polarizations, as considered in
\cite{Leggett} - \cite{Branciard}).

Then, in Sec.~\ref{sec-3}, we consider a particular quantum system
involving two entangled polarized photons. We use quantum
mechanics to calculate the joint and non-joint expected values
entering in the corresponding basic Leggett inequalities, which
have been deduced assuming only a general realism which preserves
the arrow of time. Inserting all the calculated quantum expected
values in the inequalities derived in Sec.~\ref{sec-2}, we show
how the latter are violated for certain quantum states and some
ranges of the polarizer angles involved. This means that, when
assuming the common sense concerning the arrow of time, realism,
local or not local, is incompatible with quantum mechanics. In
other words, quantum mechanics cannot be completed with any (local
or non-local) hidden variables. We stress the differences among
our conclusion and the results from the recent papers quoted
above, and remark the interest of looking for a polarized
entangled system of two photons that, like the one considered in
the present paper, leads to a violation of Leggett inequalities,
while fulfilling the requirements of experimental testing.

Finally, in Sec.~\ref{sec-4} we consider the Bohm-like hidden
variable theories \cite{Bohm} and we show why it is not obvious,
from the very beginning, that the realism of this theory can be
consistent with all experiments. Thus, at the end, the
incompatibility that we find here between realism and quantum
mechanics becomes also an incompatibility with Bohm theory itself,
against the expectations raised in \cite{Suarez} of still leaving
an open door to this theory, or against the uncritical confidence
put on it in \cite{Leggett-2}.

\section{The basic Leggett inequalities}
\label{sec-2}

Following \cite{Gr创oblacher}, let a source $S$ emit pairs of
photons with entangled polarization directions, in a given global
quantum state, $\Psi_{AB}$, towards the corresponding analyzers 1
and 2, placed, respectively, at the localities $A$ and $B$, with
orientations given by real unit vectors $\vec{a}$ and $\vec{b}$.
Notice the difference with \cite{Leggett} - \cite{Branciard},
where there is a source which emits not a well given polarized
entangled global state, $\Psi_{AB}$, as in our case, but mixtures
of different entangled polarizations according to some given
probability distribution.

Let us go to our case and assume general (local or non-local)
realism. Then, when each photon of each pair is detected, the
results of the polarization measurements are given, respectively,
by functions $A(\vec{a}, \vec{b}; \lambda)$ and
$B(\vec{b},\vec{a};\lambda)$ which, at detection, take the values
+1 or -1. Here, $\lambda$ is a supplementary (``hidden'') variable
taking values in a real domain $\Lambda$, such that, for the
ensemble of pairs of polarized entangled photons in the state
$\Psi_{AB}$, has a probability distribution $\rho_{\Psi}(\lambda)$
obeying

\begin{equation}
\rho_{\Psi}(\lambda)\geq 0 \  \  \ and \ \ \ \int_\Lambda d\lambda
\rho_{\Psi}(\lambda)=1. \label{rho}
\end{equation}

Then one has the following three expected values over the
ensemble:

\begin{equation}
\bar{A}=\int_\Lambda d\lambda
\rho_{\Psi}(\lambda)A(\vec{a},\vec{b};\lambda), \label{<A>}
\end{equation}

\begin{equation}
\bar{B}=\int_\Lambda d\lambda
\rho_{\Psi}(\lambda)B(\vec{b},\vec{a};\lambda), \label{<B>}
\end{equation}

and

\begin{equation}
\overline{AB}=\int_\Lambda d\lambda
\rho_{\Psi}(\lambda)A(\vec{a},\vec{b};\lambda)B(\vec{b},\vec{a};\lambda).
\label{<AB>}
\end{equation}

In principle, $\bar{A}$, $\bar{B}$ and $\overline{AB}$ depend on
$\Psi$ and on the set of variables $(\vec{a},\vec{b})$. Non
locality is allowed by the possible dependence of $A$ on $\vec{b}$
and of $B$ on $\vec{a}$, and realism is represented by the
supplementary variable $\lambda$.

Since the quantities $A$ and $B$ only take the values $\pm {1}$,
one has

\begin{equation}
1-\vert A-B \vert=AB=-1+\vert A+B \vert.
\label{ABequality}
\end{equation}

Then, averaging on the different values of $\lambda$, and using
the obvious inequality $\int_\Lambda d\lambda \rho_\Psi |A-B|\geq
|\int_\Lambda d\lambda \rho_\Psi (A-B)|$, one obtains the basic
Leggett inequalities

\begin{equation}
1-\vert\int_\Lambda d\lambda
\rho_{\Psi}(\lambda)(A(\vec{a},\vec{b};\lambda)-B(\vec{b},\vec{a};\lambda))\vert
\geq \int_\Lambda d\lambda
\rho_{\Psi}(\lambda)A(\vec{a},\vec{b};\lambda)B(\vec{b},\vec{a};\lambda)$$
$$\geq -1+\vert\int_\Lambda d\lambda
\rho_{\Psi}(\lambda)(A(\vec{a},\vec{b};\lambda)+B(\vec{b},\vec{a};\lambda))\vert,
\label{Leggett}
\end{equation}
or in a more compact notation
\begin{equation}
1-|\bar{A}-\bar{B}|\geq \overline{AB}\geq -1+|\bar{A}+\bar{B}|.
\label{Leggett2}
\end{equation}

It must be stressed that, unlike what it happens with the proof of
Bell's inequalities \cite{Bell}\cite{Clauser}, one does not need
the assumption of local realism in order to derive the above
Leggett inequalities. It has been enough to assume realism as
such, local or non-local. The reason for this is that now we
perform all measurements along the same directions $\vec a$ and
$\vec b$. Furthermore, each time, both measurements, along $\vec
a$ and $\vec b$, respectively, are jointly performed. This is why
we have been able to use a unique probability distribution
$\rho_{\Psi}(\lambda)$ in the calculation of the mean value of the
equations (\ref{ABequality}).

    However, notice that the above reasoning cannot lead to
the conclusion (\ref{Leggett2}) unless we assume, as we have done,
the `outcome independence' assumption \cite{Jarrett}, that is,
unless we assume that $A$ does not depend on $B$ and reciprocally.
Actually, let us write $A=A(\vec a, \vec b; \lambda, B)$ and
$B=B(\vec a, \vec b; \lambda, A)$, such that these conditions on
$A$ and $B$ do not define two functions $A=A(\vec a, \vec b;
\lambda)$ and $B=B(\vec a, \vec b; \lambda)$. That is, given the
arguments $\vec a$, $\vec b$ and $\lambda$, assume that more than
a couple of $A$ and $B$ values can exist that satisfy the
conditions $A=A(\vec a, \vec b; \lambda, B)$ and $B=B(\vec a, \vec
b; \lambda, A)$. In this case, the above probability distribution,
$\rho_\Psi$, could depend of the different couples of $A$ and $B$
values, i. e., the four couples of values $(1,1)$, $(1,-1$,
$(-1,1)$ and $(-1,-1)$. Then, we could not be sure that we can use
a unique probability distribution, $\rho_\Psi$, in order to
conclude (\ref{Leggett2}) from (\ref{ABequality}). This could be
an example of absence of a `common probability space', as it has
been argued in \cite{Hess} on different grounds.

    Nevertheless, it can be easily seen that the above outcome dependence
can only be maintained if we give up the arrow of time: in plain
words, only if we accept that the future could affect the past.
Actually, if we preserve the arrow of time, and assume, for
example, that measurements at $A$ always precede measurements at
$B$, we could have at most $A=A(\vec a, \vec b; \lambda)$ and
$B=B(\vec a, \vec b; \lambda, A)$. But then, by inserting
$A=A(\vec a, \vec b; \lambda)$ in $B=B(\vec a, \vec b; \lambda,
A)$, we finally have $B$ as a function of $\vec a$, $\vec b$ and
$\lambda$, that is, we are in fact in the original case of outcome
independence.

\section{Violation of the inequalities by quantum mechanics}
\label{sec-3}

Let us consider the quantum system of two entangled polarized
photons whose state, $\Psi_{AB}$, is
\begin{equation}
\Psi_{AB}=(1-c^2)^{1/2} u_Au_B+c v_Av_B, \label{state}
\end{equation}
where $u$ and $v$ stand for the entangled states (the kets) of two
entangled photons. The states refer to two linear polarizations in
the directions given by two unit orthogonal 3-space vectors
$\vec{u}$ and $\vec{v}$, respectively. The coefficient $c$ is a
real positive quantity such that $1\geq c$. The $A$ and $B$ index
denote the corresponding two separated entangled localities.

As explained in Sec.~\ref{sec-2}, we place at the localities $A$
and $B$ the polarization analyzers whose orientations are given by
the unit vectors $\vec a$ and $\vec b$. In the present Section, we
easily show how the basic Leggett inequalities (\ref{Leggett2})
are violated for some configurations of the vectors $\vec a$ and
$\vec b$, for some values of the coefficient $c$, according to QM,
i. e., when we insert in these inequalities the expecting values,
$\bar A$, $\bar B$ and $\overline{AB}$, dictated by QM.

Let us consider the corresponding expected values $\bar A$, $\bar
B$ and $\overline{AB}$ which appear in the basic Legget
inequalities (\ref{Leggett2}).

To begin with, we will have
\begin{equation}
\bar A = 2P_A(a,+)-1,
\label{<A>2}
\end{equation}
where $P_A(a,+)$ stands for the probability of one of the two
polarization outcomes at $A$, the one to which we have
conventionally assigned the value +1. In an analogous way we have
\begin{equation}
\bar B = 2P_B(b,+)-1, \label{<B>2}
\end{equation}
where the meaning of $P_B(b,+)$ is now obvious. To obtain
(\ref{<A>2}) and (\ref{<B>2}) we have used the completeness
relations $P_A(a,+)+P_A(a,-)=1, P_B(b,+)+P_B(b,-)=1$, where the
meaning of the non defined terms should be obvious.

For the expected value $\overline{AB}$ we will have
\begin{equation}
\overline{AB}=2[P_{AB}(a,b,++)+P_{AB}(a,b,--)]-1,\label{<AB>2}
\end{equation}

Here $P_{AB}(a,b,++)$ stands for the joint probability of having
the same kind of polarization outcome (the one to which it has
been conventionally assigned the value +1) at both analyzers.
Similarly for $P_{AB}(a,b,--)$. When getting (\ref{<AB>2}), we
have used the completeness relation
$P_{AB}(a,b,++)+P_{AB}(a,b,--)+P_{AB}(a,b,+-)+P_{AB}(a,b,-+)=1$,
where again the notation should be obvious.

Inserting the above expressions, (\ref{<A>2}), (\ref{<B>2}) and
(\ref{<AB>2}), for $\bar A$, $\bar B$ and $\overline{AB}$, in the
left hand side of the Leggett's inequalities (\ref{Leggett2}), we
find the equivalent inequality
\begin{equation}
1\geq {|P_A(a,+)-P_B(b,+)|+P_{AB}(a,b,++)+P_{AB}(a,b,--)]}.
\label{Leggett3}
\end{equation}

So, let us calculate the different probabilities which appear in
this version of the basic Leggett inequalities. To begin with, we
have:
\begin{equation}
P_A(a,+)=|\Psi_{AB}.a_A|^2,
\label{probabilityA}
\end{equation}
where $\Psi_{AB}.a_A$ stands for the Hilbert scalar product of the
kets $\Psi_{AB}$ and $a_A$. The last one refers to a polarized
photon which is at the place $A$, with the linear polarization
corresponding to $\vec a$. We easily find
\begin{equation}
P_A(a,+)=(1-c^2)\cos^2 \alpha +c^2\sin^2 \alpha,
\label{probabilityA2}
\end{equation}
where $\alpha$ is the angle between the vectors $\vec u$ and $\vec
a$.

Similarly:
\begin{equation}
P_B(b,+)= (1-c^2)\cos^2 \beta +c^2\sin^2 \beta,
\label{probabilityB2}
\end{equation}
with $\beta$ the angle between the vectors $\vec u$ and $\vec b$.

For the first joint probability, $P_{AB}(a,b,++)$, we easily find
\begin{equation}
P_{AB}(a,b,++)=|\Psi_{AB}.a_Ab_B|^2= (1-c^2)\cos^2 \alpha \cos^2
\beta+c^2\sin^2 \alpha \sin^2 \beta+ \frac{1}{2} c(1-c^2)^{1/2}
\sin 2\alpha \sin 2\beta. \label{probabilityAB+}
\end{equation}

For the other joint probability $P_{AB}(a,b,--)$, we must
calculate the expression $|\Psi_{AB}.a_{\perp A}b_{\perp B}|^2$,
where the kets $a_{\perp A}$ and $b_{\perp B}$ refer,
respectively, to two unit vectors, $\vec{a_{\perp}}$ and
$\vec{b_{\perp}}$, which are orthogonal to $\vec a$ and $\vec b$,
respectively. The result is
\begin{equation}
P_{AB}(a,b,--)=(1-c^2)\sin^2 \alpha \sin^2 \beta+c^2\cos^2 \alpha
\cos^2 \beta+ \frac{1}{2} c(1-c^2)^{1/2}\sin 2\alpha \sin 2\beta.
\label{probabilityAB-}
\end{equation}

Inserting the above probabilities (\ref{probabilityA2}),
(\ref{probabilityB2}), (\ref{probabilityAB+}) and
\ref{probabilityAB-}), in (\ref{Leggett3}), the left hand side of
the basic Leggett inequality becomes
\begin{equation}
1\geq |1-2c^2||\cos^2 \alpha -\cos^2 \beta|+\cos^2 \alpha \cos^2
\beta + \sin^2 \alpha \sin^2\beta +c(1-c^2)^{1/2} \sin 2\alpha
\sin 2\beta. \label{Leggett4}
\end{equation}

Furthermore, for $\alpha =\epsilon^{1/2}$ and $\beta=\pi /2
-\epsilon^{1/2}$, where $\epsilon$ is a real positive
infinitesimal quantity, the above inequality reduces to
\begin{equation}
1\geq |1-2c^2|(1-2\epsilon)+2\epsilon+4c(1-c^2)^{1/2}\epsilon,
\label{Leggett5}
\end{equation}
to first order in $\epsilon$.

Now, let us assume that $1>2c^2$. Hence, (\ref{Leggett5})
becomes
\begin{equation}
c\geq 2c\epsilon+2(1-c^2)^{1/2}\epsilon, \label{final eq.}
\end{equation}
to first order in $\epsilon$. Then, it is straight forward to see
that this inequality becomes slightly violated for any value of
$c$ of order $\epsilon$ but such that $c<2\epsilon$. Notice that
this infinitesimal value for $c$ fulfills the initial condition
$1>2c^2$. In other words, the basic Leggett inequalities
(\ref{Leggett2}) are in contradiction with QM. But, as we have
already noticed at the end of Sec.~\ref{sec-2}, these inequalities
are deduced by only assuming any sort of (local or non-local)
realism plus the arrow of time (in particular, we do not use the
the crypto non-local realism assumption, used in \cite{Leggett} -
\cite{Branciard}). Then, as it has been announced in the
Introduction, if the arrow of time is preserved, QM and realism,
local or non-local, are incompatible: QM cannot be
\emph{completed}, i. e., the given quantum state and the
measurement `direction' cannot be supplemented with any, local or
non local, hidden variables in order to complete the statistical
predictions of the QM with sure predictions for individual
measurements, unless we allow for an unphysical violation of the
arrow of time.

As we have mentioned in the Introduction, in \cite{Branciard2}
some Legget inequalities have been deduced for \emph{well definite
polarizations} too, under the assumption of crypto non-local
realism . These inequalities are experimentally violated when the
inserted joint expected value, $\overline{AB}$, is the one
calculated according to QM, for an entangled pair of two polarized
photons in the negative parity state, i. e., in the singlet state.
Nevertheless, in our case, where we assume no other condition that
realism, local or non-local, jointly with the arrow of time, plus
QM (that is plus the expected values $\bar{A}$, $\bar{B}$ and
$\overline{AB}$, dictated by QM), it can be seen that there is no
contradiction between the corresponding Leggett inequalities
applied to the above singlet state and QM (likewise, there is
neither contradiction for the positive parity state, or for the
singlet state of two entangled 1/2 spin particles: the details of
the calculation will be given elsewhere). This is why we have
considered a state like state (\ref{state}), which is more general
that the ones considered above for entangled photons, in order to
display the above contradiction between realism and QM. Now, in
the framework of a certain realism, the assumption of crypto
non-local realism may sound reasonable when we have a source of
entangled polarized photons which \emph {produces a mixture of
different polarizations}, according to some probability
distribution. This is the case in\cite{Leggett} -
\cite{Branciard}. Nevertheless, if we consider that the source
produces definite polarizations, we must leave the crypto
non-local realism hypothesis and simply accept the testable
quantum expected values for $\bar A$ and $\bar B$. This is just
what we have done in the present Section.

Since in the present paper we only want to prove that QM, on one
hand, and realism as such plus the arrow of time preservation, on
the other hand, are incompatible, at the end we have chosen a very
special case of the more general quantum state (\ref{state}), in
order to accomplish the proof in the most simple way.
Nevertheless, the importance of the result deserves that some
quantum state be found (perhaps the same considered in
Sec.~\ref{sec-3}) such that, while still leading to the violation
of the basic Leggett inequalities (\ref{Leggett2}), be able to
fulfill the requirements of an experimental test.

\section{Non-local realism and Bohm theory. Conclusions}
\label{sec-4}

It seems at first sight that saying that QM cannot be completed,
as we say, should be erroneous since it has been largely claimed
that a hidden variable theory (HVT) exists, the Bohm's theory
\cite{Bohm}, which is, at the same time, a non local realistic
theory and one which reproduces all the predictions of QM.
Furthermore, the kind of realism assumed in Bohm's theory seems to
be the most general kind of realism one can conceive. To begin
with, in this HVT theory, the assumed hidden variables "depend
both on the state of the measuring apparatus and the observed
system" (\emph{contextual} realism). On the other hand, even if
implicitly, Bohm only places his hidden variables behind any
\emph{actually} obtained outcome measurement (\emph{actual}
realism), and not behind a merely obtainable outcome (\emph{
joint} realism) \cite{Lapiedra}. This kind of contextual actual
realism is also the one considered here, for example when proving
(\ref{Leggett2}), and the one implicitly considered in
\cite{Leggett} - \cite{Branciard}.

Now, let us consider in detail whether is it true that Bohm's HVT
is always consistent with this \emph{actual}, apparatus dependent,
non local realism. Bohm proves that his theory gives the same
probability of finding a particle in a given position that QM
does. From this, he reasonably concludes that his "interpretation
is capable of leading in all possible experiments to identical
predictions to those obtained from the usual interpretation", that
is to say, from those obtained from QM. Then, when considering an
entangled extended system, as in Einstein-Podolsky-Rosen
experiments (similar to the ones considered by Bell in his seminal
papers), Bohm assumes that his realism is non local. In this way,
his non local HVT can explain the observed violation of the
ordinary Bell inequalities, in agreement with QM, without having
to give up realism (see \cite{Bell} for example).

But is it always this way? Is it true that we can devise actual
non-local HVT that lead to the same predictions that QM, \emph{
for all conceivable experiments}? Let us make some considerations
in order to show why, from the very beginning, it is not evident
that HVT can always agree with QM and at the same time with non
local realism.

First of all, in these theories, each time that one performs a
measurement on the particle position, if one wants to complete,
beyond the obtained outcome, the precedent particle trajectory
with a new trajectory piece, one must provide the probability
density of the particle position just after this outcome. The
provided probability becomes the new initial probability. Then,
this initial probability must be taken the same as the one
dictated by standard QM if we want the HVT to agree henceforth
with QM. After this, in the HVT framework, one does not need to
worry about how this initial probability evolves in time until one
performs a subsequent measurement, since HVT are just designed to
predict the same probability evolution as the predicted by
Schr\"odinger equation. Hence, when some consecutive different
measurements are performed on the same particle \cite{Lapiedra},
or better on the different particles of an entangled system, one
expects to find some well definite correlations among the
corresponding outcomes: the correlations dictated by QM and
observed in Bell type experiments.

More precisely: let it be a system of two entangled polarized
photons. Assume, for mere sake of simplicity, that we measure both
polarizations at a simultaneous time, i. e., both measurements are
space-like events. Either in QM or in HVT, the probabilities of
the two simultaneous measurements outcomes are given by the
corresponding initial quantum entangled state, just the previous
one to both measurements: in the notation of the precedent
sections, states $u$ and $v$ conveniently entangled. In HVT, these
initial quantum states are supplemented with the (uncritically)
assumed initial values of some non-local hidden variables,
$\lambda$ (actually, $\lambda$ plus direction $\vec b$, for
example, in the notation of Sec.~\ref{sec-2}), whose deterministic
time evolution, in absence of measurement, preserves, as it must
be, the quantum evolution of the outcome probabilities. This
evolution of the $\lambda$ values can always be established and
this is the great triumph of Bohm theory. Nevertheless, the point
here is that this evolution of $\lambda$, which mimics so
perfectly well the quantum evolution of the above probabilities,
has nothing to do with the explanation of the quantum
correlations, as for instance the ones which are behind the
reported quantum violation of inequalities (\ref{Leggett2}). It
has nothing to do \emph{since these correlations have only to do
with the initial $\lambda$ values}, which are uncritically assumed
to exist, plus the quantum state entangling $u$ and $v$, which
actually exist. Then, is it sure from the very beginning that
these correlations will always be compatible with some non local
realism, that is, with the assumption that some initial values of
non local hidden variables, $\lambda$, are behind all these
outcome probabilities? No, we cannot be sure of this
compatibility, unless we be able to prove it. But, as we have
seen, HVT, though uncritically assuming it, do not actually prove
it, while the result in the last Section saying that QM cannot be
completed can be seen as a counter example showing that this
cannot be proved, since such non local hidden variables do not
always exist. It is true nevertheless that, according to the end
of Sec.~\ref{sec-2}, one could still argue that a consistent Bohm
theory could exist by allowing it to violate the arrow of time.
But such a strange addition to a full realistic theory, like Bohm
theory, would really be an unnatural addition.

Thus, it seems that there is no room left "for models that force
Nature to mimic the concept of trajectory" as it is still expected
in \cite{Suarez}.

To summarize: according to the above discussions, either Quantum
Mechanics, or a realism that preserves the arrow of time, must be
false. So, if on the ground of its general success we accept QM,
plus the arrow of time, we must conclude that realism as such, i.
e., local or nonlocal, should contradict experiments, an statement
that would deserve being tested. Then the answer to the Leggett
question \cite{Leggett-2} of "it is indeed realism rather than
locality which has to be sacrificed?" would be `yes'. All in all:
against Einstein's old dream, it seems that QM cannot be
completed.

\begin{acknowledgements}
This work was partially supported by the project PAPIIT IN113607,
DGAPA-UNAM, M\'exico (M. S) and by the Spanisnh Ministerio de
Educaci\'on y Ciencia, MEC-FEDER project FIS2006-06062 (R. L.). M.
S. thanks for hospitality at the Facultad de Astronom\'\i a y
Astrof\'\i sica de la Universidad de Valencia, Spain, where part
of this work was performed. R. L. recognizes fruitful discussions
with Eugenio  Roldan and Arcadi Santamaria.
\end{acknowledgements}

%e-mails:

%Ramon.Lapiedra@uv.es, socolovs@nucleares.unam.mx,
%socolovs@iafe.uba.ar

\end{document}